\documentclass[10pt]{iopart}
\usepackage{bm,graphicx, color}

\begin{document}

\title{Dynamical membrane curvature instability controlled by intermonolayer friction}
\author{Anne-Florence Bitbol$^1$, Jean-Baptiste Fournier$^1$, Miglena I. Angelova$^{1,2}$ and Nicolas Puff$^{1,2}$}
\address{$^1$Laboratoire Mati\`ere et Syst\`emes Complexes (MSC),
Universit\'e Paris Diderot, Paris 7 and UMR CNRS 7057, 10 rue Alice Domon et L\'eonie Duquet, F-75205 Paris Cedex 13, France}
\address{$^2$ Universit\'e Pierre et Marie Curie, Paris 6, 4 place Jussieu, F-75252 Paris Cedex 05, France}
\ead{miglena.anguelova@upmc.fr}

\begin{abstract}
We study a dynamical curvature instability caused by a local chemical modification of a phospholipid membrane. In our experiments, a basic solution is microinjected close to a giant unilamellar vesicle, which induces a local chemical modification of some lipids in the external monolayer of the membrane. This modification causes a local deformation of the vesicle, which then relaxes. We present a theoretical description of this instability, taking into account both the change of the equilibrium lipid density and the change of the membrane spontaneous curvature induced by the chemical modification. We show that these two types of changes of the membrane properties yield different dynamics. In contrast, it is impossible to distinguish them when studying the equilibrium shape of a vesicle submitted to a global modification. In our model, the longest relaxation timescale is related to the intermonolayer friction, which plays an important part when there is a change of the equilibrium density in one monolayer. We compare our experimental results to the predictions of our model by fitting the measured time evolution of the deformation height to the solution of our dynamical equations. We obtain a good agreement between theory and experiments. Our fits enable us to estimate the intermonolayer friction coefficient, yielding values consistent with previous measurements.
\end{abstract}

\pacs{87.16.dj, 87.10.-e, 68.03.Cd}
\submitto{\JPCM}

\maketitle

\section{Introduction}

Much more than simple barriers, membranes play an active role in many biological phenomena, in particular when they bend and curve into a multitude of shapes~\cite{McMahon05,Zimmerberg06}. Indeed, each membrane shape is coupled to a specific function, and the diversity and dynamics of these shapes are vital for cell physiology~\cite{Farsad03,Shnyrova09}. The remarkable material properties of biological membranes, inherent in their nature of complex bi-dimensional visco-elastic films, play an important part in membrane deformations.

Giant unilamellar vesicles (GUVs) are model membranes made of amphiphilic lipid molecules that self-assemble into closed bilayers in water. Their sizes and curvatures are similar to those of living cells and their lipid bilayer membrane exhibits the basic properties of biological membranes. These features have made them very attractive objects to study the physics of a number of phenomena in cellular biology. Although lacking membrane proteins and a cytoskeleton, they have been used as a minimal cell model to mimic various biological processes such as membrane budding and endocytosis, fusion and fission, transport phenomena across the membrane, lipid domain formation, etc.~\cite{Cans03,Streicher09,Campelo08,Baumgart03,Yanagisawa07,Veatch02}. Especially, giant vesicle ``micromanipulation'' gives a unique opportunity to carry out controlled experiments on an individual vesicle, obtaining data \textit{directly} while exposing this vesicle to mechanical, biochemical, or chemical perturbations~\cite{Wick96,Angelova99,Borghi03,Khalifat08}. In this last case, a localized chemical gradient is created close to the membrane by microinjection. The response of the membrane to such a localized gradient can be significantly different from the one caused by a uniform perturbation of its environment. In particular, we will show in this paper that a change of the spontaneous curvature and a change of the preferred area per lipid lead to different dynamics in the case of a local perturbation. In contrast, when uniform perturbations are studied, observing the change of the equilibrium shape does not enable to distinguish between these two types of changes of the membrane properties~\cite{Lee99}.

In a previous work~\cite{Fournier09}, we have reported both theoretically and experimentally a chemically-driven membrane shape instability observed when GUVs are submitted to a local pH increase. In this instability, a local chemical modification affecting some lipids in one monolayer of the membrane changes the preferred area per lipid, thereby inducing a transient local bilayer curvature. Since only one monolayer is affected, the lateral redistribution of the lipids is strongly slowed down by the intermonolayer friction. When such a chemical modification is applied to a small enough surface, it triggers the ejection of a tubule growing exponentially toward the chemical source.

In this paper, we describe in more detail the first phase of this local curvature instability, before the tubule ejection, and we present a more complete comparison of our experimental results with the predictions of our model. In the experiments, the chemical modification of the vesicle membrane has been achieved by locally delivering to the membrane outer leaflet a basic solution of NaOH with a micropipette. We have performed a time-step chemical modifications (\textquotedblleft pulse\textquotedblright experiments) and we have measured the time evolution of the height of the deformation of the vesicle with respect to its initial shape. Our theoretical model has been completed by taking into account in our dynamical equations the spontaneous curvature change in addition to the equilibrium density change induced by the chemical modification. We have also compared these two effects. The comparison of the experimental results with the predictions of our model has been performed by fitting the measured deformation height with the solution of the dynamical equations of our model. Several ``pulse'' experiments, done on three different GUVs, have been analyzed. The agreement between the experiments and the model is quite good, and our fits enable us to estimate the intermonolayer friction coefficient, yielding values consistent with the literature.

\section{Materials and Methods}

\subsection{Membrane composition and giant vesicle preparation}

The following lipids were used without further purification: egg yolk L-$\alpha$-phosphatidylcholine (EYPC), Sigma, Lyon, France; Brain L-$\alpha$-phosphatidylserine (PS), Avanti Polar Lipids, Alabaster, AL. All others chemicals were of highest purity grade: HEPES, Interchim, Montlu\c{c}on, France; EDTA, Sigma; NaOH, Sigma. 

Giant vesicles were formed by the liposome electroformation method~\cite{Angelova86} in a thermostated chamber. The particular electroformation protocol used in this work was the following: lipid mixture solutions were prepared in chloroform/diethyl ether/methanol (2:7:1) with a total lipid concentration of 1 mg/ml. All the liposome preparations were made with a unique lipid mixture of EYPC and PS with EYPC/PS 90:10 mol/mol. A droplet of this lipid solution (1 $\mu$l) was deposited on each of the two parallel platinum wires constituting the electroformation electrodes, and dried under vacuum for 15 min. An AC electrical field, 10 Hz, 0.26 Vpp, was applied to the electrodes. Buffer solution (2 ml, pH 7.4, HEPES 0.5 mM, EDTA 0.5 mM, temperature $25^\circ\mathrm{C}$) was added to the working chamber (avoiding agitation). The voltage was gradually increased (for more than two hours) up to 1 Vpp and maintained during 15 more minutes, before switching the AC field off. The GUV were then ready for further use. In each preparation at least 10 GUVs of diameter 50-80 $\mu$m were available.

\subsection{Microscopy imaging and micromanipulation}

We used a Zeiss Axiovert 200M microscope, equipped with a charged-coupled device camera (CoolSNAP HQ; Photometrics, Tucson, AZ). The experiments were computer-controlled using the Metamorph software (Molecular Devices, Downington, PA). The morphological transformations and the dynamics of the membrane were followed by phase contrast microscopy.

Tapered micropipettes for the local injection of NaOH were made from GDC-1 borosilicate capillaries (Narishige, Tokyo, Japan), pulled on a PC-10 pipette puller (Narishige, Tokyo, Japan). The inner diameter of the microcapillary used for performing the local injections onto a GUV was $0.3~\mu\mathrm{m}$. For these local injections, a microinjection system (Eppendorf femtojet) was used. The micropipettes were filled with a basic solution of NaOH ($1~\mathrm{M}$, pH~$13$). The injected volumes were on the order of picoliters, the injection lasted a few seconds, and the injection pressure was 100 hPa. The positioning of the micropipettes was controlled by a high-graduation micromanipulator (MWO-202; Narishige, Tokyo, Japan). The injections were performed at different distances from the GUV surface, taking care to avoid any contact with the lipid membrane. The injected solution covered about 10 \% of the GUV surface at the time of the injection. Estimations based on visualizing the flux from the micropipette and taking into account the dilution of NaOH in the GUV formation buffer (pH 7.4) yield, at the deformation onset, a pH ranging from 8 to 9 on the vesicle membrane.

\section{Experiments}
\label{section_exp}
Giant unilamellar vesicles in the fluid phase were formed by electroformation at $25^\circ\mathrm{C}$ in a buffer at pH~$7.4$ from a mixture of EYPC and PS with EYPC/PS 90:10 mol/mol. 

The chemical modification of the membrane was achieved by locally delivering to the membrane outer leaflet a basic solution of NaOH ($1~\mathrm{M}$, pH~$13$). This local increase of the pH should affect the head groups of the phospholipids PS and PC forming the membrane. 
Indeed, the amino group of the PS head group deprotonates at high pH, its intrinsic $\mathrm{pK_a}$ being about 9.8 in vesicles constituted of a PC/PS 90:10 mol/mol mixture~\cite{Tsui86}. Besides, the positively charged trimethylammonium group of the PC head group associates with hydroxide ions at high pH, the dissociation constant K of this equilibrium being such that $\mathrm{pK}=14-\mathrm{pK_a^{eff}}$, where $\mathrm{pK_a^{eff}}=11$~\cite{Lee99}. Both reactions increase the negative charge of the lipids, which entails a local change of the preferred area per lipid head group in the outer leaflet.

\begin{figure}[h t b]
\centering
\includegraphics[width=0.75\textwidth]{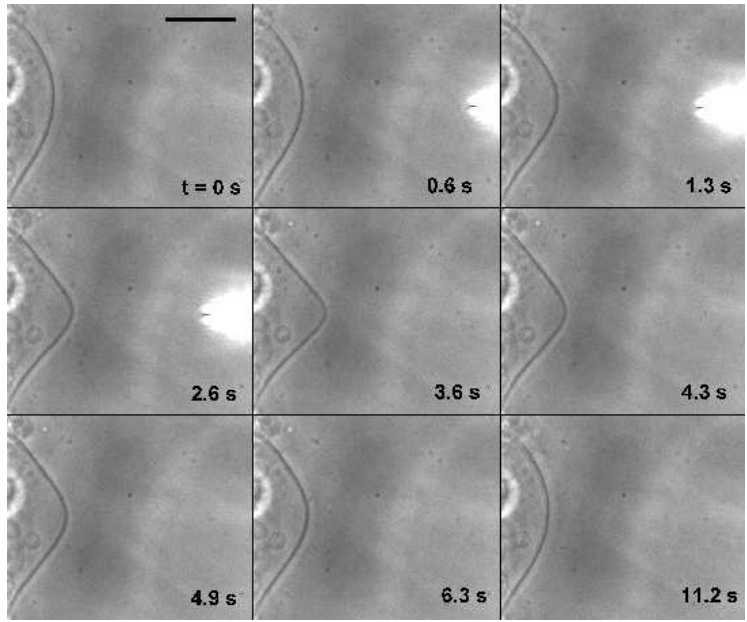}
\caption[]{Time-step chemical modification of the vesicle membrane. A local modulation of the pH at the level of a vesicle membrane induces a smooth deformation of the vesicle (frames 1.3 to 3.6 s). The deformation is completely reversible when the NaOH delivery is stopped (frames 4.3 s to the end). Scale bar: 50 $\mu\mathrm{m}$.
\label{Fig_planche}}
\end{figure}

Figure~\ref{Fig_planche} shows a typical \textquotedblleft pulse\textquotedblright experiment. We try to perform a time-step chemical modification. For this, we quickly approach the pipette without injecting any solution, and then we inject the solution during a time $\Delta t\approx\!3~\mathrm{s}$ before quickly withdrawing the pipette. 
One can see in Fig.~\ref{Fig_planche} the vesicle before any microinjection (frame 0 s), and the micropipette during its approach (frames 0.6 and 1.3 s). A two-step process occurs. First, a smooth deformation of the vesicle starts to develop toward the pipette, i.e., opposite to the flow (frames 0.6 to 3.6 s). At this point we stop the injection and we quickly withdraw the pipette, and the membrane deformation relaxes after reaching a maximum (frames 4.3 s to the end). 

The following control experiments have been carried out: (i) We have checked that no deformation occurs if only buffer solution is injected. (ii) In order to verify that the observed effects were not simply due to charge screening and/or osmotic effects, a local injection of salt solution (NaCl instead of NaOH) has been performed. The typical smooth and reversible deformation has not been observed in these control experiments, which shows that the pH increase is crucial in our instability.

\begin{figure}[h t b]
\centering
\includegraphics[width=0.55\textwidth]{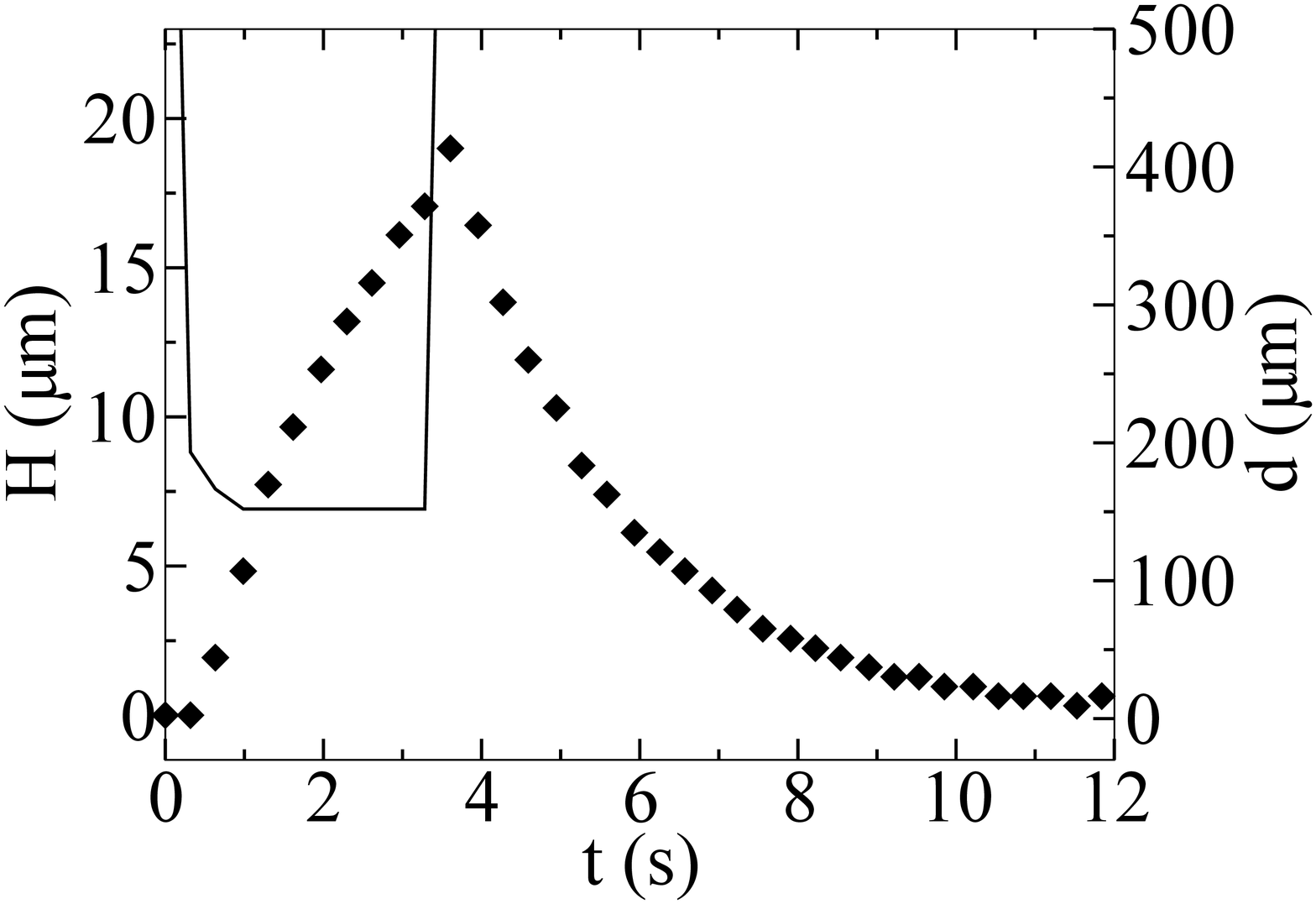}
\caption[]{Typical example of the time evolution of the vesicle deformation, measured in front of the pipette, in a \textquotedblleft pulse\textquotedblright  experiment. Dots: time evolution $H(t)$ of the deformation amplitude. Solid line: Time evolution $d(t)$ of the distance of the micropipette from the electrode supporting the GUV. \label{Fig_H_d}}
\end{figure}

Fig.~\ref{Fig_H_d} shows the measured time evolution of the height $H(t)$ of the deformation of the vesicle with respect to its initial shape for the experiment presented in Fig.~\ref{Fig_planche}. 
One can see easily on this figure the two-step process described previously. For more clarity, the distance of the micropipette from the electrode supporting the GUV is also presented. 

Several \textquotedblleft pulse\textquotedblright experiments were conducted on three different GUVs. The complete analysis of these experiments is presented in Sec.~\ref{Fit}.

\section{Theoretical model}
\subsection{Free energy of a bilayer membrane}
The free energy per unit area $f$ of a lipid membrane is well described on a large scale by
\begin{equation}
f=\sigma_0+\frac{\kappa}{2}c^2-\kappa c_0^b c\,.
\label{f}
\end{equation}
This free-energy density depends on the membrane curvature $c$, which is defined as the sum of the two principal curvatures on the bilayer midsurface. The constitutive constants $\sigma_0$, $\kappa$ and $c_0^b$ denote, respectively, the membrane tension, its bending elastic constant, and its spontaneous curvature.
The resulting membrane free energy
\begin{equation}
F=\int dA\,f\,
\label{intaire}
\end{equation}
is known as the Helfrich Hamiltonian~\cite{Helfrich73}. The integral in Eq.~(\ref{intaire}) is a surface integral over the area $A$ of the bilayer midsurface. Note that we do not include any Gaussian curvature term in this free energy. Indeed, the topology of the vesicle is not affected by the instability we wish to study, so that the integral of the Gaussian curvature, and thus its contribution to $F$, remains constant by virtue of the Gauss-Bonnet theorem~\cite{Buchin}.

To account for the bilayer structure of the membrane, its free-energy density $f$ can be written as the sum of the free-energy densities of the two monolayers, which will be noted $f^+$ and $f^-$. Since the curvature $c$ is defined on the bilayer midsurface, it is the same for both monolayers. Furthermore, we assume that the two monolayers have the same lipid composition before the onset of the instability, so that they have identical tensions and bending rigidities. They also have opposite spontaneous curvatures, noted $\pm c_0$, since the lipids in the two monolayers are oriented in opposite directions: their hydrophilic heads are oriented towards the exterior of the bilayer, while their hydrophobic tails are oriented towards the interior of the bilayer. To study our instability, it is necessary to take into account the inhomogeneities in the lipid mass densities $\rho^\pm$ defined on the bilayer midsurface in each monolayer. Defining the scaled densities $r^\pm=(\rho^\pm-\rho_0)/\rho_0$, where $\rho_0$ is a reference density, which is chosen identical for both monolayers, we may write~\cite{Seifert93}
\begin{equation}
f^\pm=\frac{\sigma_0}{2}+\frac{\kappa}{4}c^2\pm\frac{\kappa c_0}{2}c+\frac{k}{2}\left(r^\pm \pm ec\right)^2\,.
\label{fpm}
\end{equation}
In this formula, $k$ is the stretching elastic constant of a monolayer, which is the same for both monolayers as they are identical, and $e$ denotes the distance between the neutral surfaces of the monolayers~\cite{Safran} and the midsurface of the bilayer. Indeed, the scaled densities in each monolayer at a distance $e$ from the bilayer midsurface are $r_n^\pm=r^\pm \pm ec$ at first order in the small variable $ec$, so that if $f^\pm$ is written in terms of the curvature and $r_n^\pm=(\rho_n^\pm-\rho_0)/\rho_0$, these two variables are decoupled. Such a decoupling between deformations where only the curvature is modified (bending) and deformations in which only the density is affected (stretching) is characteristic of the neutral surface~\cite{Safran}. We choose the sign convention for the curvature in such a way that a spherical vesicle has $c<0$. Then the monolayer denoted ``$+$" in Eq.~(\ref{fpm}) is the outer monolayer.

The expression (\ref{fpm}) of the free-energy density of a monolayer is a general second-order expansion around a reference state characterized by a flat shape ($c=0$) and a uniform density $\rho^\pm=\rho_0$ \cite{Futur}. It is valid for small deformations around this reference state: $r^\pm=\mathcal{O}(\epsilon)$ and $ec=\mathcal{O}(\epsilon)$, where $\epsilon$ is a small nondimensional parameter characterizing the deformation.

If the densities on the neutral surfaces, $\rho_n^\pm$, are both equal to the reference value $\rho_0$, summing the monolayer free-energy densities $f^\pm$ in (\ref{fpm}) gives back the standard free-energy density (\ref{f}) of the bilayer: $f=f^+ +f^-$. Note that we find $c_0^b=0$ for the spontaneous curvature of the bilayer, since we have considered two identical monolayers.

\subsection{Modification of the free energy due to a local pH change}
\label{modif}

In the experiment, when the pipette expels the NaOH solution close to the GUV, it induces a local increase of the pH, which affects the head groups of the phospholipids forming the membrane, as explained in Sec.~\ref{section_exp}. As some of these head groups become more negatively charged during this modification, the preferred area per lipid head group increases locally. Pursuing the analysis of Ref.~\cite{Fournier09}, we call $\phi(\mathbf{r},t)$ the fraction of the lipids of the external monolayer that are chemically modified by the local pH increase in the experiment. Since the pH on the membrane should never exceed 9, this fraction should remain very small (see the $\mathrm{pK_a}$ values in Sec.~\ref{section_exp}), so we may assume $\phi=\mathcal{O}(\epsilon)$. 

Since the characteristic timescale of an acido-basic reaction is determined by the diffusion time of the reactants~\cite{Eigen64}, we can consider that $\phi(\mathbf{r},t)$ follows instantaneously the pH field outside the vesicle. The hydroxide ions diffuse in the buffer solution with a diffusion coefficient $D_{OH^-}\approx5\times10^3\,\mu\mathrm{m^2/s}$ \cite{Daniele99}. Thus, in the typical $5\,\mathrm{s}$ of relaxation of the membrane deformation, they diffuse on a length of approximatively $1.5 \times 10^2\,\mu\mathrm{m}$. This length is larger than the width of the instability and comparable with the size of the GUV. Besides, the pH field is probably also affected by the advection created by the flux coming from the micropipette, then by its retraction and last by the movement of the membrane. However, the central zone of the instability remains in contact with the zone of highest pH. At present, we have no knowledge of this time-dependent pH field beyond these rough estimates. Hence, to simplify, we are going to assume that $\phi(\mathbf{r},t)$ does not evolve significantly during the time of the instability. More precisely, we make the simplifying hypothesis that $\phi(\mathbf{r})$ is time-independent after the end of the injection ($t=0$):
\begin{quote}
\emph{(SH$_1$)}\quad $\phi(\mathbf{r},t)=\phi(\mathbf{r})$ for $t\ge0$.
\end{quote}
We also assume that the inner monolayer is not affected by the pH increase outside the vesicle. The permeation coefficient of hydroxide ions through a lipid membrane is $P_{OH^-}\approx 10^{-3}-10^{-5}\,\mathrm{cm/s}$, which yields a negligible pH increase inside the vesicle on the timescale of our experiments \cite{Elamrani83, Seigneuret86}.

\emph{A priori}, the constitutive constants of monolayer ``$+$" are all affected by the chemical modification, i.e., they depend on $\phi(\mathbf{r})$. However, since we focus on small deformations around the flat shape and the uniform density $\rho^+=\rho_0$, we can write to second order in $\epsilon$:
\begin{eqnarray}
f^+&=&\frac{\sigma_0}{2}+\sigma_1\phi+\frac{\sigma_2}{2}\phi^2+\tilde\sigma\left(1+r^+\right)\phi\ln\phi
+\frac{\kappa}{4}c^2+\frac{\kappa}{2}\left(c_0+\tilde c_0\phi\right)c\nonumber\\&+&\frac{k}{2}\left(r^++ ec\right)^2\,.
\label{fmod}
\end{eqnarray}
With respect to expression~(\ref{fpm}), the monolayer tension and spontaneous curvature of monolayer ``$+$" have been modified according to 
\begin{eqnarray}
\frac{\sigma_0}{2}&\to&\frac{\sigma_0}{2}+\sigma_1\phi+\frac{\sigma_2}{2}\phi^2+\tilde\sigma(1+r^+)\phi\ln\phi\,,\\
c_0&\to&c_0+\tilde c_0\phi\,.
\end{eqnarray}
The change of the other constitutive constants ($k$, $\kappa$ and $e$) is irrelevant as far as the free-energy density at second order is concerned. We have assumed that the constitutive constants were analytical functions of $\phi$, apart from the $\tilde\sigma(1+r^+)\phi\ln\phi$ term, which corresponds to a mixing entropy term per unit area \cite{Futur}. Except from this additional term, expression~(\ref{fmod}) is the most general second-order expansion of the monolayer free-energy density (in the three small variables $ec$, $r^+$ and $\phi$) when the total masses of modified and non-modified lipids remain constant \cite{Futur}.

\subsection{Force density in the membrane}

The force density in each monolayer of a bilayer with lipid density and composition inhomogeneities described by the free-energy densities~(\ref{fpm})--(\ref{fmod}) has been derived in Ref.~\cite{Futur}. It reads to first order in $\epsilon$
\begin{eqnarray}
p_i^+&=&-k\,\partial_i\left(r^++ec-\frac{\sigma_1}{k}\phi\right)\,,\label{pip}\\
p_i^-&=&-k\,\partial_i\left(r^--ec\right)\,,\label{pim}\\
p_n&=&\sigma_0 c-\tilde{\kappa}\Delta c-k e\,\Delta \left(r^+-r^-\right) - \frac{\kappa \tilde{c}_0}{2}\Delta\phi\nonumber\\
 &=&\sigma_0 c-\tilde{\kappa}\Delta c-k e\,\Delta \left(r^+-r^--\frac{\sigma_1}{k}\phi\right) - \frac{\kappa \bar{c}_0}{2}\Delta\phi\label{pn}\,,
\end{eqnarray}
where $p_i^\pm$ is the force density in monolayer ``$\pm$" acting in a direction $i$ tangential to the membrane, while $p_n=p_n^++p_n^-$ is the total normal force density in the membrane. In these formulas, we have defined the constants $\tilde\kappa=\kappa+2ke^2$ and $\bar c_0=\tilde c_0+2\sigma_1 e/\kappa$. The symbol $\partial_i$ denotes partial derivation in the direction $i$, while $\Delta$ is the covariant Laplacian operator.

Let us discuss the terms depending on $\phi$ in the force densities. For a given composition $\phi(\mathbf{r})$, the equilibrium density of a plane monolayer with fixed total mass is obtained by minimizing the free energy per unit mass $f^\pm/\rho^\pm$ for $c=0$. Considering a fixed total mass is justified here since the timescales of our instability are much shorter than the flip-flop characteristic time, which is assumed not to be significantly modified by the local chemical modification we study. 
For monolayer ``$+$", this gives at first order in $\epsilon$
\begin{equation}
r^+_{\mathrm{eq}}(\phi)\equiv \frac{\rho^+_\mathrm{eq}(\phi)-\rho_0}{\rho_0}=\frac{\sigma_0/2+\sigma_1\phi}{k}\,,
\label{req}
\end{equation}
where we have used the fact that $\sigma_0\ll k$. Thus, the chemical modification changes the scaled equilibrium density of the plane monolayer ``$+$" by the amount
\begin{equation}
\delta r^+_\mathrm{eq}=r^+_\mathrm{eq}(\phi)-r^+_\mathrm{eq}(0)=\frac{\sigma_1}{k}\phi\,.
\label{dreq}
\end{equation}
We observe that this change of the equilibrium density at $c=0$ appears in the force densities (\ref{pip}) and (\ref{pn}), which is in agreement with the simpler theory of Ref.~\cite{Fournier09}. 
The equilibrium density obtained by minimization at $c=0$, i.e. for a flat membrane, will be referred to as the ``plane-shape equilibrium density" in the following.

Let us now focus on the term  $-\frac{1}{2}\kappa \bar{c}_0\Delta\phi$ in (\ref{pn}), which comes from the chemical modification too, but which is not related to the change of the plane-shape equilibrium density. The membrane is at mechanical equilibrium if the force density vanishes. It can be seen from (\ref{pip}) and (\ref{pn}) that the flat shape $c=0$ is a solution to $\bm{p}=\mathbf{0}$ at a given inhomogeneous $\phi(\mathbf{r})$ only if $\bar c_0=0$. Thus, $\bar c_0\phi$ (and not $\tilde c_0\phi$) represents the actual change of the spontaneous curvature of monolayer ``$+$" caused by the chemical modification.

\subsection{Change of the spontaneous curvature and of the equilibrium density}
\label{Mod_subsec}
Changing the local spontaneous curvature of a bilayer membrane may result in shape or budding instabilities~\cite{Tsafrir01,Tsafrir03,Staneva05}. Alternatively, affecting locally the plane-shape equilibrium density non-symmetrically in each monolayer of a bilayer can also yield shape or budding instabilities~\cite{Sens04, Fournier09}. However, until now, no study has considered an asymmetric modification of both the spontaneous curvature and the plane-shape equilibrium density in the two monolayers. As can be seen from Eq.~(\ref{pn}), changing the plane-shape equilibrium density of the lipids produces a destabilizing normal force per membrane unit area $\delta p_n^{(1)}=e\sigma_1\Delta\phi$, while changing the spontaneous curvature yields $\delta p_n^{(2)}=-\frac{1}{2}\kappa \bar c_0\Delta\phi$. Thus, both of these changes should induce a shape or budding instability.

In the case where only the equilibrium density is affected, which corresponds to $\sigma_1\ne0$ and $\bar c_0=0$, the budding should vanish as soon as the lipid density has relaxed to its new equilibrium value, even if the modified lipids remain in place. Indeed, if $r^-$ is homogeneous, and if $r^+(\mathbf{r})$ reaches $r_\mathrm{eq}^+(\mathbf{r})$ (defined in Eq.~(\ref{req})), the equilibrium condition $p_n=p_i^\pm=0$ is satisfied for $c=0$, which means that the flat shape is an equilibrium shape. On the contrary, in the case where only the spontaneous curvature is modified, i.e., assuming $\sigma_1=0$ and $\bar c_0\ne0$, the budding persists as long as the modified lipids remain in place. Indeed, the equilibrium condition $p_n=p_i^\pm=0$ can be satisfied only if $c(\mathbf{r})$ verifies $\sigma_0 c(\mathbf{r})-\kappa\Delta c(\mathbf{r})=\frac{1}{2}\kappa\bar c_0\Delta\phi(\mathbf{r})$, which implies that $c(\mathbf{r})\ne0$ if $\Delta\phi(\mathbf{r})\ne0$, so that the plane shape is not an equilibrium shape for a generic inhomogeneous $\phi(\mathbf{r})$. Thus, although both mechanisms should lead to a shape or budding instability, they are not physically equivalent. We will study the differences between them in more detail in Sec.~\ref{comp}.

It is interesting to determine the relative importance of the two above-mentioned mechanisms in a budding instability. For this, let us first compare the relative variation of the spontaneous curvature $d c_0/c_0$ and the relative variation of the plane-shape equilibrium density $d\rho_\mathrm{eq}^+/\rho_\mathrm{eq}^+$ induced by the chemical modification in monolayer $+$. As explained in Sec.~\ref{section_exp}, this chemical modification affects the head groups of some lipids, which become more negatively charged, so that the preferred area of the head group of these molecules increases. This means that our chemical modification may be viewed roughly as an increase of the preferred area per lipid head group. It is now necessary to resort to a microscopic model to express the variation of the spontaneous curvature and of the plane-shape equilibrium density as a function of the variation of the preferred area per lipid head group. 

\begin{figure}[h t b]
\centering
\includegraphics[width=0.3\textwidth]{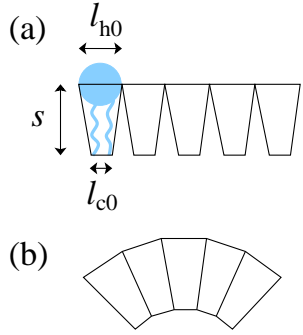}
\caption[]{Two-dimensional illustration of our simple geometrical model: the lipid head groups have a characteristic size $l_\mathrm{h0}$, and the chains have a smaller characteristic size $l_\mathrm{c0}$. The lipids are viewed as solid cones favoring close-packing. (a) Plane shape: the area per lipid is given by $a_\mathrm{h0}=l_\mathrm{h0}^2$. (b) The spontaneous curvature $c_0$ of the monolayer is obtained when both the heads and the chains are close-packed. The conical shape of the lipids has been exaggerated for clarity: in reality, $c_0 s\ll 1$.\label{Figgeom}}
\end{figure}

The simplest way of doing this is to take a mere geometrical model in which each lipid is constituted of a head group with area $a_\mathrm{h0}=l_\mathrm{h0}^2$ and of a chain group with a smaller area $a_\mathrm{c0}=l_\mathrm{c0}^2$, situated at a fixed distance $s$ from each other (see Fig.~\ref{Figgeom}). Both the head and the chain are supposed to be incompressible, and to favor close-packing. In this crude model, the plane-shape equilibrium density is equal to $m/a_\mathrm{h0}$, where $m$ is the mass of a lipid (see Fig.~\ref{Figgeom} (a)). The spontaneous curvature is obtained when  both the heads and the chains are close-packed, i.e. when the area per chain is equal to $a_\mathrm{c0}$ and the area per head is equal to $a_\mathrm{h0}$ (see Fig.~\ref{Figgeom} (b)): geometry yields $c_0=2(l_\mathrm{h0}-l_\mathrm{c0})/(l_\mathrm{c0} s)$. If the preferred area per lipid head group $a_\mathrm{h0}$, or equivalently the corresponding characteristic length $l_\mathrm{h0}$ is modified, it gives $\delta\rho_\mathrm{eq}/\rho_\mathrm{eq}=-2 \delta l_\mathrm{h0}/l_\mathrm{h0}$ and $\delta c_0/c_0=2 \delta l_\mathrm{h0}/(l_\mathrm{c0} c_0 s)$. Thus, given that $c_0 s\gg1$ (since lipids spontaneously organize into planar membranes and not micelles), we obtain
\begin{equation}
\frac{\delta c_0}{c_0}\approx -\frac{1}{c_0 s} \frac{\delta\rho_\mathrm{eq}}{\rho_\mathrm{eq}}\,.
\label{geom}
\end{equation}

A less crude microscopic model which we may use to express the variation of $c_0$ and $\rho_\mathrm{eq}$ caused by a change of the preferred area per head group $a_\mathrm{h0}$ was exposed in Ref.~\cite{Miao94}. In this model, the head and chain of a lipid have respective preferred areas $a_\mathrm{h0}$ and $a_\mathrm{c0}$, while in the monolayer, the actual areas of these groups are $a_\mathrm{h}$ and $a_\mathrm{c}$. The elastic energy per molecule is written as 
\begin{equation}
g(a_\mathrm{h},a_\mathrm{c})=\frac{K_\mathrm{h}}{2}a_\mathrm{h0}\left(\frac{a_\mathrm{h}}{a_\mathrm{h0}}-1\right)^2+\frac{K_\mathrm{c}}{2}a_\mathrm{c0}\left(\frac{a_\mathrm{c}}{a_\mathrm{c0}}-1\right)^2\,,
\label{modeleMiao}
\end{equation}
where $K_\mathrm{h}$ and $K_\mathrm{c}$ are stretching elastic constants for each group. In Ref.~\cite{Miao94}, it has been shown that this free energy per molecule can be used to derive the area-difference elasticity model of a membrane. Following the same lines, it is possible to show that Eq.~(\ref{modeleMiao}) yields the following free energy per unit area of a monolayer:
\begin{equation}
 f^\pm=\frac{\kappa}{4}c^2\pm \frac{\kappa c_0}{2}c+\frac{k}{2}\left (r^\pm \pm ec\right)^2\,,
\end{equation}
which corresponds to our free-energy density Eq.~(\ref{fpm}) in the case where the reference density $\rho_0$ is taken equal to $\rho_\mathrm{eq}$ (which implies $\sigma_0=0$ \cite{Futur}). The calculations in Ref.~\cite{Miao94} provide the following expressions of $c_0$ and $\rho_\mathrm{eq}$ from microscopic constants: 
\begin{eqnarray}
c_0&=&\frac{(K_\mathrm{h}a_\mathrm{c0}+K_\mathrm{c}a_\mathrm{h0})(a_\mathrm{h0}-a_\mathrm{c0})}{s(K_\mathrm{h}+K_\mathrm{c})a_\mathrm{h0}a_\mathrm{c0}}\,,\\
\rho_\mathrm{eq}&=&\frac{m}{a_0}=m\frac{K_\mathrm{h}a_\mathrm{c0}+K_\mathrm{c}a_\mathrm{h0}}{(K_\mathrm{h}+K_\mathrm{c})a_\mathrm{h0}a_\mathrm{c0}}\,,
\end{eqnarray}
where $a_0$ is the equilibrium area per lipid on the neutral surface of the monolayer. It is straightforward to express the relative variations $\delta c_0/c_0$ and $\delta\rho_\mathrm{eq}/\rho_\mathrm{eq}$ as a function of the relative variation of the preferred area per head group $\delta a_\mathrm{h0}/a_\mathrm{h0}$, which yields
\begin{equation}
\frac{\delta c_0}{c_0}=-\frac{K_\mathrm{h}a_\mathrm{c0}^2+K_\mathrm{c}a_\mathrm{h0}^2}{K_\mathrm{h}a_\mathrm{c0}(a_\mathrm{h0}-a_\mathrm{c0})}\,\frac{\delta\rho_\mathrm{eq}}{\rho_\mathrm{eq}}\approx-\frac{K_\mathrm{h}+K_\mathrm{c}}{K_\mathrm{h}}\,\frac{1}{c_0 s}\,\frac{\delta\rho_\mathrm{eq}}{\rho_\mathrm{eq}}\,,
\label{Miao}
\end{equation}
where we have used the fact that $c_0 s=(a_\mathrm{h0}-a_\mathrm{c0})/a_0\ll1$. The values of the elastic constants $K_\mathrm{h}$ and $K_\mathrm{c}$ should be of the same order, and we may hint at $K_\mathrm{c}<K_\mathrm{h}$ since the chains are able to reorganize spatially. The result (\ref{geom}) from the crude geometrical model is recovered when $K_\mathrm{h}\to \infty$ while $K_\mathrm{c}$ is finite, as it should.
Since $s\approx e$, the results (\ref{geom}) and (\ref{Miao}) coming from the two microscopic models can both be written in the form
\begin{equation}
\left|\frac{\delta c_0}{c_0}\right|\approx \frac{\alpha}{e c_0} \left|\frac{\delta\rho_\mathrm{eq}}{\rho_\mathrm{eq}}\right|\,,
\label{oom}
\end{equation}
where the order of magnitude of the numerical coefficient $\alpha$ is one. 

We may now compare the destabilizing pressures $\delta p_n^{(1)}$ and $\delta p_n^{(2)}$ caused by each effect:  $|\delta p_n^{(2)}|/|\delta p_n^{(1)}|=(\frac{1}{2}\kappa|\bar c_0|)/(2e|\sigma_1|)$. The relative variation of the plane-shape equilibrium density in monolayer ``$+$'' induced by the chemical modification is $\delta\rho_\mathrm{eq}^+/\rho_0= \delta r_\mathrm{eq}^+=\sigma_1\phi/k$ (see Eq.~(\ref{dreq})). Besides, the relative variation of the spontaneous curvature caused by the same chemical modification is $\delta c_0/c_0=\bar c_0\phi/c_0$. Using Eq.~(\ref{oom}) and the former expressions yields $|\bar c_0|/|\sigma_1|\approx\alpha/(ek)$. Therefore, we obtain $|\delta p_n^{(2)}|/|\delta p_n^{(1)}|=\alpha\kappa/(2ke^2)$. Using the well-known orders of magnitudes $\kappa\approx10^{-19}\,\mathrm{J}$, $e\approx1\,\mathrm{nm}$ and $k\approx0.1\,\mathrm{J/m^2}$~\cite{Safran}, we find out that the two destabilizing pressures should have the same order of magnitude.

Thus, in the normal force density Eq.~(\ref{pn}), the destabilizing term $\delta p_n^{(1)}$ due to the change of plane-shape equilibrium density should usually be comparable to the other destabilizing term $\delta p_n^{(2)}$ which comes from the change of the spontaneous curvature. We are thus going to take both of these effects into account in our hydrodynamic description of the instability.

\subsection{Hydrodynamic equations}
\label{hydro}
As we focus on small deformations with respect to the plane shape, it is convenient to describe the membrane in the Monge gauge, i.e., by its height $z=h(x,y)$ with respect to a reference plane, $x$ and $y$ being Cartesian coordinates in the reference plane. Then, $c=\nabla^2 h+\mathcal{O}(\epsilon^2)$ for small deformations such that $\partial_i h=\mathcal{O}(\epsilon)$ and $\partial_i\partial_j h=\mathcal{O}(\epsilon)$ where $i,j\in\{x,y\}$. Let us denote by $\mathbf{v}^\pm(x,y,t)$ the two-dimensional velocities of the lipids within the monolayers. Recall that the fraction of chemically modified lipids, $\phi(x,y)$, is assumed to be time-independent, being fixed by the surrounding pH field modelled as a static field ($SH_1$). Since we have assumed $\phi=\mathcal{O}(\epsilon)$, and since the flow is induced by the chemical modification, we also have $\mathbf{v}=\mathcal{O}(\epsilon)$. In the Monge gauge, the force densities (\ref{pip})--(\ref{pn}) become \cite{Futur}
\begin{eqnarray}
p_i^+(x,y)&=&-k\,\partial_i\left(r^++e\nabla^2 h-\frac{\sigma_1}{k}\phi\right)\,,\label{pip_b}\\
p_i^-(x,y)&=&-k\,\partial_i\left(r^--e\nabla^2 h\right)\,,\label{pim_b}\\
p_z(x,y)&=&\sigma_0 \nabla^2 h-\tilde{\kappa}\nabla^4 h-k e\,\nabla^2 \left(r^+-r^-\right)- \frac{\kappa \tilde{c}_0}{2}\nabla^2\phi\label{pn_b}\,.
\end{eqnarray}

The fraction of the chemically modified lipids may be expanded in Fourier modes: $\phi(x,y)=\sum_\mathbf{q}\phi_\mathbf{q}\,e^{i(q_x x+q_y y)}$. Since we work at first order in $\epsilon$, the hydrodynamic equations are going to be linear, which justifies the use of a Fourier expansion. Since the region where the lipids are modified has a well-defined width $\sim\!q^{-1}$ (see Sec.~\ref{Fit}), we are going to identify $\phi(x,y)$ with its dominant Fourier mode. 
This approximation will be discussed in Sec.~\ref{Sec_discussion}. 
Furthermore, we will consider a wavevector $\mathbf{q}$ parallel to the $x$ axis (this can be done without any loss of generality by choosing the orientation of the axes appropriately). We thus make a second (and last) simplifying hypothesis:
\begin{quote}
\emph{(SH$_2$)}\quad
One-mode approximation: $\phi(x,y)\approx\phi_q\,e^{iqx}$.
\end{quote}
Following this hypothesis, we may also write $r^\pm(x,y,t)=r^\pm_q(t)\,e^{iqx}$, $v_x^\pm(x,y,t)= v_q^\pm(t)\,e^{iqx}$ and $h(x,y,t)= h_q(t)\,e^{iqx}$ at linear order in $\epsilon$.

The dynamical equations describing the joint evolution of the membrane shape and of the lipid densities have been developed in another context in Refs.~\cite{Seifert93,Evans94}. They have been applied to the study of the instability caused by a local density change in the case where $\bar c_0=0$ in Ref.~\cite{Fournier09}. In terms of the above Fourier components, they read
\begin{eqnarray}
&&\hspace{-1cm}
-\eta_2q^2v_q^+
-ikq\left(r_q^+ - e q^2 h_q - \frac{\sigma_1}{k}\phi_q\right)
-2\eta q v_q^+
-b\left(v_q^+ - v_q^-\right)=0\,,
\label{un}\\
&&\hspace{-1cm}
-\eta_2q^2v_q^-
-ikq\left(r_q^- + e q^2 h_q\right)
-2\eta q v_q^-
+b\left(v_q^+ - v_q^-\right)=0\,,
\label{deux}\\
&&\hspace{-1cm}
-\left(\sigma_0 q^2+\tilde\kappa q^4\right)h_q
+k e q^2\left(r_q^+-r_q^-\right)
+ \frac{\kappa \tilde{c}_0}{2}q^2\phi_q
-4\eta q\frac{\partial h_q}{\partial t}=0\,,
\label{trois}\\
&&\hspace{-1cm}
\frac{\partial r_q^\pm}{\partial t} + iq v_q^\pm=0\,.
\label{quatre}
\end{eqnarray}
Equations~(\ref{un}) and (\ref{deux}) are generalized Stokes equations describing the balance of the forces per unit area acting tangentially in monolayer ``$+$" and in monolayer ``$-$", respectively. The first term in each of these equations corresponds to the viscous force density due to the two-dimensional flow of the lipids, $\eta_2$ being the two-dimensional viscosity of the lipids. The second term is the density of elastic forces given by Eqs.~(\ref{pip_b}) and (\ref{pim_b}). The third term corresponds to the viscous stress exerted by the flow of the surrounding fluid, which is set in motion by the two-dimensional flow of the lipids in the membrane. The viscosity of the external fluid is denoted by $\eta$. The last term is the stress originating from the intermonolayer friction~\cite{Evans94}, $b$ being the intermonolayer friction coefficient. Equation~(\ref{trois}) describes the balance of the forces per unit area acting normally to the membrane. Its first three terms represent the elastic force density given by Eq.~(\ref{pn_b}). Its last term is the normal viscous stress exerted by the flow of the surrounding fluid, the normal velocity of which matches $\partial h/\partial t$ on the membrane. Finally, Equation~(\ref{quatre}) expresses the conservation of mass at first order in $\epsilon$. Typical values of the dynamical parameters, used throughout, are $\eta_2\approx10^{-9}\,\mathrm{J\,s/m^2}$, $\eta\approx10^{-3}\,\mathrm{J\,s/m^3}$, and $b\approx10^8-10^9\,\mathrm{J\,s/m^4}$~\cite{Pott02,Shkulipa06}.

Comparing these dynamical equations with the ones of Ref.~\cite{Fournier09} shows that $-\sigma_1 \phi_q/k$  corresponds to the scalar field $\epsilon(\mathbf{r},t)$ introduced in Ref.~\cite{Fournier09} to describe the change of the equilibrium density, which is in agreement with  Eq.~(\ref{dreq}). Ref.~\cite{Fournier09} focused on the effect of the variation of the plane-shape equilibrium density, which corresponds to taking $\bar c_0=0$ here. Here, both the change of the equilibrium density and the change of the spontaneous curvature are taken into account.

\subsection{Resolution of the hydrodynamic equations}
\label{resol}
Let us define the (symmetric) average scaled density $\bar r_q(t)=r_q^-+r_q^+$ and the (antisymmetric) differential scaled density $\hat r_q(t)=r_q^+-r_q^-$. Eliminating $v_q^\pm$ in
Eqs.~(\ref{un})--(\ref{deux})  thanks to Eq.~(\ref{quatre}) and adding them together gives
\begin{equation}
\frac{\partial\bar r_q}{\partial t}=-\frac{kq}{\eta_2q+2\eta}
\left(\bar r_q-\displaystyle\frac{\sigma_1}{k}\phi_q\right)\,,
\label{dynabar}
\end{equation}
while substracting them yields
\begin{equation}
\hspace{-1.2cm}
\frac{\partial}{\partial t}\left(\begin{array}{c}
q h_q\\\\
\hat r_q
\end{array}\right)=-
\left(\begin{array}{cc}
\displaystyle\frac{\sigma_0 q+\tilde\kappa q^3}{4\eta}
&-\displaystyle\frac{keq^2}{4\eta}\\\\
-\displaystyle\frac{keq^3}{b}
&\displaystyle\frac{kq^2}{2b}
\end{array}\right)
\left(\begin{array}{c}
\displaystyle q h_q\\\\
\hat r_q
\end{array}\right)
+\left(\begin{array}{c}
\displaystyle\frac{\kappa \tilde c_0 q^2}{8\eta}\phi_q\\\\
\displaystyle\frac{\sigma_1 q^2}{2 b}\phi_q
\end{array}\right),
\label{relax}
\end{equation}
where we have assumed $\eta_2q^2\ll b$ and $\eta q\ll b$, which is well verified for $\pi/q\approx40\,\mathrm{\mu m}$ corresponding to the width of the deformation observed in the experiments.

Equation~(\ref{dynabar}) shows that $\bar r_q$ relaxes to its equilibrium value $\sigma_1\phi_q/k$ with a very short timescale $\tau=-(\eta_2q+2\eta)/(kq)$. Indeed, with typically $\pi/q\approx40\,\mathrm{\mu m}$, and with the parameters given above, we obtain $\tau\approx0.3\,\mathrm{\mu s}$, which can be considered instantaneous in our experiment. 

Equation~(\ref{relax}) can be solved by diagonalizing the square matrix involved. In our experiments, we have $q\ll\sqrt{\sigma_0/\tilde\kappa}$. Indeed, the tension of a vesicle is superior to about $10^{-7} \,\mathrm{J/m^{2}}$, so $\sqrt{\sigma_0/\tilde\kappa}\geq 10^6 \,\mathrm{m^{-1}}$. In this regime, the eigenvalues of the square matrix in Eq.~(\ref{relax}) are 
\begin{equation}
\gamma_1\approx \frac{kq^2}{2b}\mathrm{\,\,and\,\,}\gamma_2\approx\frac{\sigma_0 q}{4\eta}\,. 
\label{eigenvalues}
\end{equation}
This result can be checked rapidly by noting that, in the regime where $q\ll\sqrt{\sigma_0/\tilde\kappa}$, the coefficient $keq^3/b$ is much smaller than all the other coefficients in the matrix, so that the square matrix in Eq.~(\ref{relax}) may be approximated by an upper triangular matrix. The deformation thus evolves according to 
\begin{equation}
h_q(t)=A e^{-\gamma_1t}+Be^{-\gamma_2t}+\frac{\kappa\bar c_0}{2\sigma_0}\phi_q\,,
\label{evol_h_gen}
\end{equation}
where the constants $A$ and $B$ can be determined from the initial conditions on $h_q$ and $\hat r_q$. The term $\kappa\bar c_0 \phi_q/(2\sigma_0)$ comes from the constant solution of the inhomogeneous equation (\ref{relax}), which corresponds physically to the residual deformation at equilibrium. We find that this residual deformation vanishes if $\bar c_0=0$, i.e., in the case where only the plane-shape equilibrium density (and not the spontaneous curvature) is modified, which is consistent with the previous discussions. The characteristic times $\gamma_1^{-1}$ and $\gamma_2^{-1}$ are both much longer than $\tau$, the longest one being $\gamma_1^{-1}$, which gives a total relaxation time of some seconds. This slow relaxation originates from the strong intermonolayer friction, which is involved in the antisymmetric changes of the monolayer densities, as the lipids of one monolayer move relative to the other monolayer.

\subsection{Qualitative picture of the dynamics in two limiting cases}
\label{comp}
In Sec.~\ref{Mod_subsec}, we have shown that it is physically different to change locally the plane-shape equilibrium density and to change locally the spontaneous curvature. Now that we have studied the dynamics of the instability, we are going to analyze the difference between these two effects. Although both of them occur in the generic case (see Sec.~\ref{Mod_subsec}), we are going to study the two limiting cases when only one of these changes occurs.

\begin{figure}[h t b]
  \begin{center}
    \includegraphics[width=.45\textwidth]{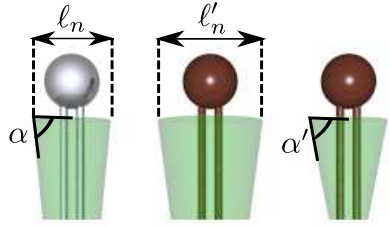}
    \caption{First lipid (light grey): non-modified lipid. The length $\ell_n$ represents the preferred diameter per lipid on the neutral surface, while the angle $\alpha$ quantifies the preferred curvature. Second and third lipids (dark grey or brown): modified lipids. For the second lipid, the modification affects only the preferred density on the neutral surface, but not the preferred curvature. It is the contrary for the third lipid.}
  \label{presentation}
  \end{center}      
\end{figure}

Let us first introduce a schematic representation of lipids, which takes into account the preferred area per lipid on the neutral surface, which is (by definition) independent of the membrane curvature, and their preferred curvature (see Fig~\ref{presentation}). The preferred shape of a lipid is represented by a cone, superimposed on the lipid. The area per lipid on the neutral surface (within the monolayer) is symbolized by the upper surface of this cone, while its preferred curvature is symbolized by the angle of this cone. 
 
\subsubsection{Dynamics in the case where $\bar c_0=0$}

We are first going to study the dynamics of the relaxation in the case where $\bar c_0=0$. The interest of studying this theoretical case is to show how modifying the plane-shape equilibrium density and not the spontaneous curvature can induce a transient shape instability.

\begin{figure}[h t b]
  \begin{center}
    \includegraphics[width=.3\textwidth]{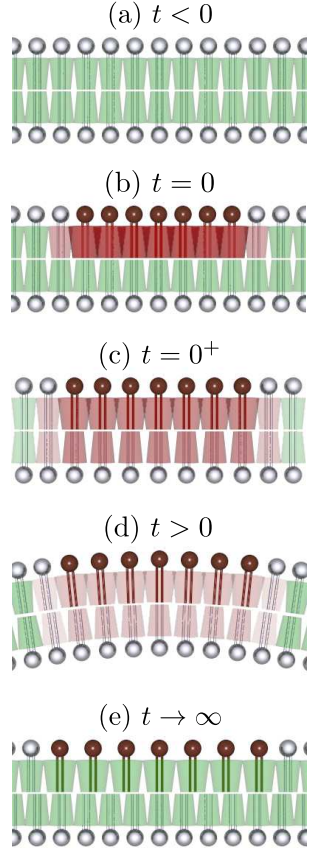}
    \caption{Qualitative description of the dynamics of the instability if $\bar c_0=0$. The lipids that become chemically modified at $t=0$ are represented in dark grey (brown). The cones are dark (red) for the lipids experiencing compression or dilation with respect to this preferred area and in light grey (green) otherwise. The intensity of the grey (red) colour represents the degree of the compression or dilation. Here, the modification affects the preferred density on the neutral surface, but not the preferred curvature.}
  \label{dessin}
  \end{center}      
\end{figure}

Let us consider simple initial conditions at $t=0$: $\bar r_q(0)=\hat r_q(0)=h_q(0)=0$. This corresponds to the idealized case where the injection of the NaOH solution is local in time: then,  following \textit{(SH$_1$)}, we have $\phi(\mathbf{r},t)=0$ for $t<0$ and $\phi(\mathbf{r},t)=\phi(\mathbf{r})$ for $t\ge0$. At $t=0$, the preferred area per lipid suddenly increases for the lipids of the external monolayer that are chemically modified, since we expect their extra negative charge to increase repulsion. Thus, these modified lipids are effectively compressed (see Fig.~\ref{dessin}(b)). 

Since the coupled dynamics of $\hat r_q$ and $h_q$ is much slower than the dynamics of $\bar r_q$, we may consider that at $t=0^+$, the equilibrium state $\bar r_q(0^+)=\sigma_1\phi_q/k$ has been reached, while $\hat r_q(0^+)=0$ still holds, so $r_q^\pm(0^+)=\frac{1}{2}\sigma_1\phi_q/k$. This means that after an infinitesimal time, half the compression of the lipids of the external monolayer is relaxed. However, the flow of the lipids of the external monolayer has dragged the lipids of the inner monolayer, because of the intermonolayer friction, thus inducing a dilation equal to the compression in the external monolayer (see Fig.~\ref{dessin}(c)). 

After this very fast first step, the time evolution of the membrane deformation is given by
\begin{equation}
h_q(t)=\phi_q\frac{eq\sigma_1}{4\eta}\frac{e^{-\gamma_2t}-e^{-\gamma_1t}}{\gamma_2-\gamma_1}\,,
\label{ci_zero}
\end{equation}
which corresponds to Eq.~(\ref{evol_h_gen}) in the case where $\bar c_0=0$ and with the above initial conditions at $t=0^+$. The deformation appears and increases until a time $t_\mathrm{max}$ when it reaches a maximum \cite{Fournier09}, before decaying exponentially with timescale $\gamma_1^{-1}$. At the same time, the differential density $\hat r_q(t)=r_q^+-r_q^-$ decreases with timescale $\gamma_1^{-1}$ (see Fig.~\ref{dessin}(d)). The membrane curving and one monolayer sliding with respect to the other are two distinct responses to the discrepancy between the equilibrium densities of the two monolayers. Here, both of these phenomena occur in the transient regime, but the final state ($t\rightarrow\infty$) corresponds to a non-deformed membrane (since $\bar c_0=0$): the discrepancy is finally solved by the relative sliding of the monolayers, which is the slowest process (see Fig.~\ref{dessin}(e)). 

\subsubsection{Dynamics in the case where $\sigma_1=0$}

\begin{figure}[h t b]
  \begin{center}
    \includegraphics[width=.3\textwidth]{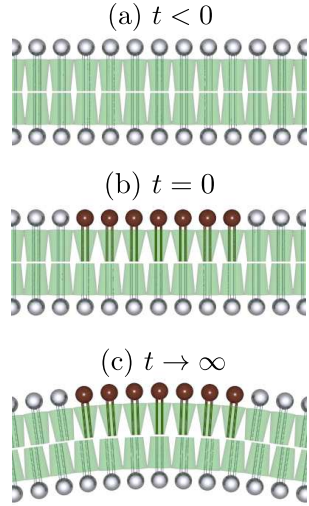}
    \caption{Qualitative description of the dynamics of the instability if $\sigma_1=0$. Here, the modification affects the preferred curvature, but not the preferred density on the neutral surface.}
  \label{dessin2}
  \end{center}      
\end{figure}

Let us now discuss the opposite case, where only the spontaneous curvature is modified. We take the same initial conditions as in the previous section: $\bar r_q(0)=\hat r_q(0)=h_q(0)=0$. At $t=0$, the preferred curvature per lipid suddenly increases for the lipids of the external monolayer that are chemically modified (see Fig.~\ref{dessin2}(b)). Here, solving Eq.~(\ref{dynabar}) shows that $\bar r_q$ remains equal to zero. If the usual approximation~(\ref{eigenvalues}) is used for the eigenvalues $\gamma_1$ and $\gamma_2$, the time evolution of the membrane deformation is given by
\begin{equation}
h_q(t)=\phi_q\frac{\kappa\bar c_0}{2\sigma_0}\left(1-e^{-\gamma_2t}\right)\,,
\label{ci_zero_2}
\end{equation}
which corresponds to Eq.~(\ref{evol_h_gen}) in the case where $\bar \sigma_1=0$ and with the above initial conditions. The deformation increases exponentially with timescale $\gamma_2^{-1}$ towards a deformed final state, where $h_q=\kappa\bar c_0\phi_q/(2\sigma_0)$ (see Fig.~\ref{dessin2}(c)). 

Note that the long timescale $\gamma_1^{-1}$ involving the intermonolayer friction does not appear in Eq.~(\ref{ci_zero_2}). Indeed, as the plane-shape equilibrium density is not modified here, the lipids should not have to slide with respect to the other monolayer. In fact, since the membrane is curved in the final state, a small sliding is necessary as the equilibrium density on the membrane midsurface is curvature-dependent, contrary to the equilibrium density on the neutral surface of each monolayer. This small sliding is visible on Fig.~\ref{dessin2}(c). This effect can be accounted for by solving our full dynamical equations (\ref{relax}), without using the approximation~(\ref{eigenvalues}) for the eigenvalues. Indeed, solving these equations for typical values of the parameters yields a small increase of the differential density $\hat r_q(t)=r_q^+-r_q^-$ with timescale $\gamma_1^{-1}$, and a small contribution to the deformation $h_q$ with timescale $\gamma_1^{-1}$. It has been checked that this contribution to the deformation is negligible.

Note that in real experimental conditions, \textit{(SH$_1$)} cannot hold indefinitely. In our experiment, this is mainly due to the diffusion of the hydroxide ions (see Sec.~\ref{modif}). Even if the modification of the lipids was irreversible, the modified lipids would diffuse in the monolayer, so the deformation would relax anyway. However, the corresponding relaxation timescale would be much longer than the previous ones given the small diffusion coefficient of the lipids in a membrane $D_\mathrm{lip}\approx1\,\,\mu\mathrm{m^2/s}$ \cite{Schmidt96}: the modified lipids would take about $L^2/D_\mathrm{lip}\approx10^4 \,\,\mathrm{s}$ to diffuse on a length $L\approx10^2\,\,\mu\mathrm{m}$.

\subsubsection{Comparison of the two types of changes of the membrane properties}
Solving the dynamical equations in the two limiting cases has confirmed that a local change of the plane-shape equilibrium density and a local change of the spontaneous curvature yield different dynamics. Indeed, for the length scales we consider, when only the plane-shape equilibrium density is modified, the deformation increases rapidly before relaxing towards a non-deformed shape with a long timescale $\gamma_1^{-1}$ (typically of several seconds). When only the spontaneous curvature is affected, the deformation increases towards a stationary curved state with a short timescale $\gamma_2^{-1}$ (typically shorter than 1 s) before relaxing slowly with a timescale of about one hour. 

In contrast, when a global modification of the environment of a vesicle is considered, studying the change of its equilibrium shape does not allow to distinguish between a change of the spontaneous curvature and a change of the preferred area per lipid \cite{Lee99}. As a matter of fact, the equilibrium shape of a vesicle with fixed volume is fully determined within the area-difference elasticity (ADE) model by the value of the combined quantity 
\begin{equation}
\overline{\Delta a_0}=\Delta a_0+\frac{2}{\alpha}c_0^b\,,
\end{equation}
where $\Delta a_0$ is the nondimensionalized preferred area difference between the two monolayers, $\alpha$ a nondimensional number involving the elastic constants of the membrane and $c_0^b$ denotes the spontaneous curvature of the bilayer \cite{Miao94}. The shape variations observed when such global modifications are performed have been interpreted as coming from a change of the spontaneous curvature $c_0^b$, under the assumption that the preferred area per lipid was not modified \cite{Lee99,Dobereiner99,Petrov99}.

Thus, the dynamical study of our instability induced by a local chemical modification of the lipids should allow to distinguish two types of changes of the membrane properties, which cannot be distinguished for a static global modification. Since both of them should be involved generically (see Sec.~\ref{Mod_subsec}), studying the dynamics may enable to determine their relative importance. However, this is difficult to do precisely in the present experiments because of the diffusion of the hydroxide ions (see Sec.~\ref{modif}). 

\section{Comparison between the experiments and the model}
\label{Fit}

\subsection{Fits of the experimental results}
\label{Fits_subsec}
In the previous section, we have exposed a theoretical model for the curvature instability observed in our experiments. In order to compare the experimental results to the predictions of this model, we are going to fit the deformation height measured during a ``pulse'' experiment with Eq.~(\ref{evol_h_gen}). This equation corresponds to the general solution of the dynamical equations of our model, and it should describe the time evolution of the height $H(t)$ of the deformation of the vesicle with respect to its initial shape. 

We present the analysis of several  ``pulse'' experiments conducted on three different GUVs (numbered 1, 2 and 3 in the following). For each experiment, we take as an initial condition the last point where the pipette is present. If we call $H_0$ the height of the deformation at this time, which will be referred to as $t=0$ as in our theoretical analysis, this initial condition can be written $H(0)=H_0$. The experimental data is thus fitted with the formula
\begin{equation}
H(t)=(H_0-C-B)e^{-\gamma_1 t}+B e^{-\gamma_2 t}+C\,,
\label{Formule_fit}
\end{equation}
which corresponds to Eq.~(\ref{evol_h_gen}) with the above initial condition. There are thus four free parameters in our fits: $B$, $C$, $\gamma_1$ and $\gamma_2$. Besides, since we expect that the diffusion of the hydroxide ions is no longer negligible after a few seconds (see Sec.~\ref{modif}), we have carried out the fits on identical intervals of duration $5\,\mathrm{s}$ for all the experiments, starting from the end of the injection ($t=0$). Thus, we eliminate the longer-term evolution which should no longer be well-described by our model. 

\begin{figure}[h t b]
\centering
\includegraphics[width=0.5\textwidth]{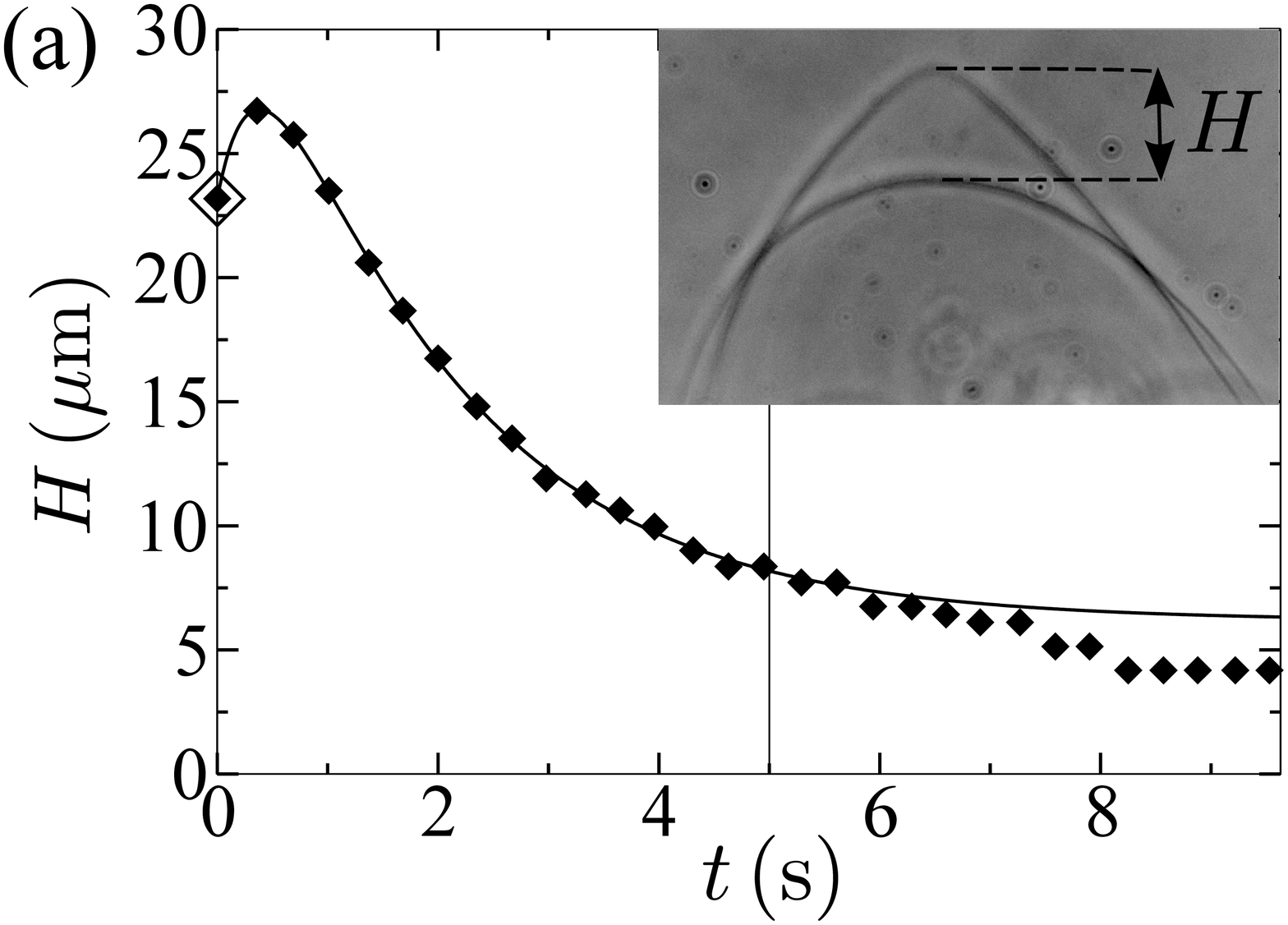}\\
\includegraphics[width=0.5\textwidth]{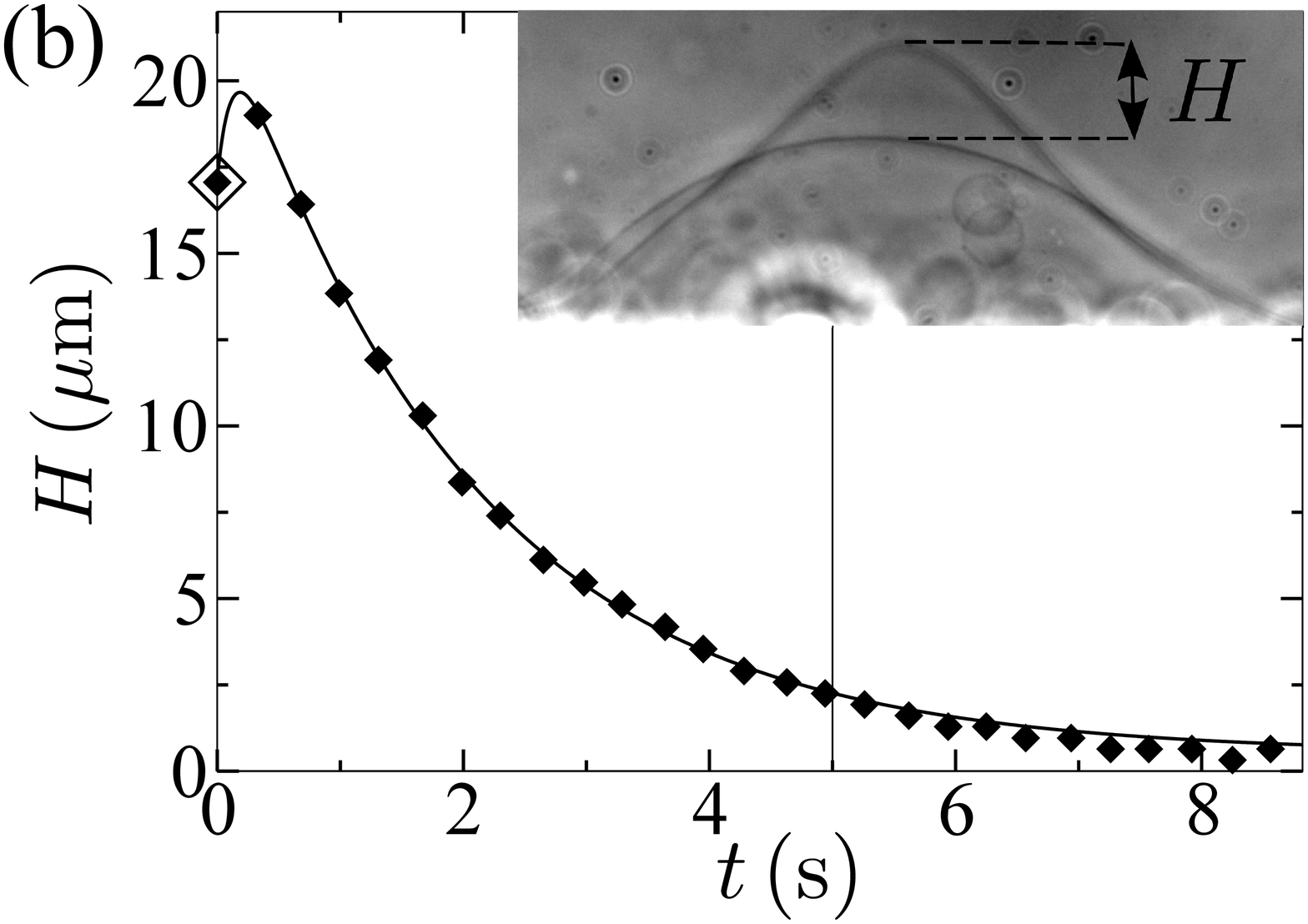}\\
\includegraphics[width=0.5\textwidth]{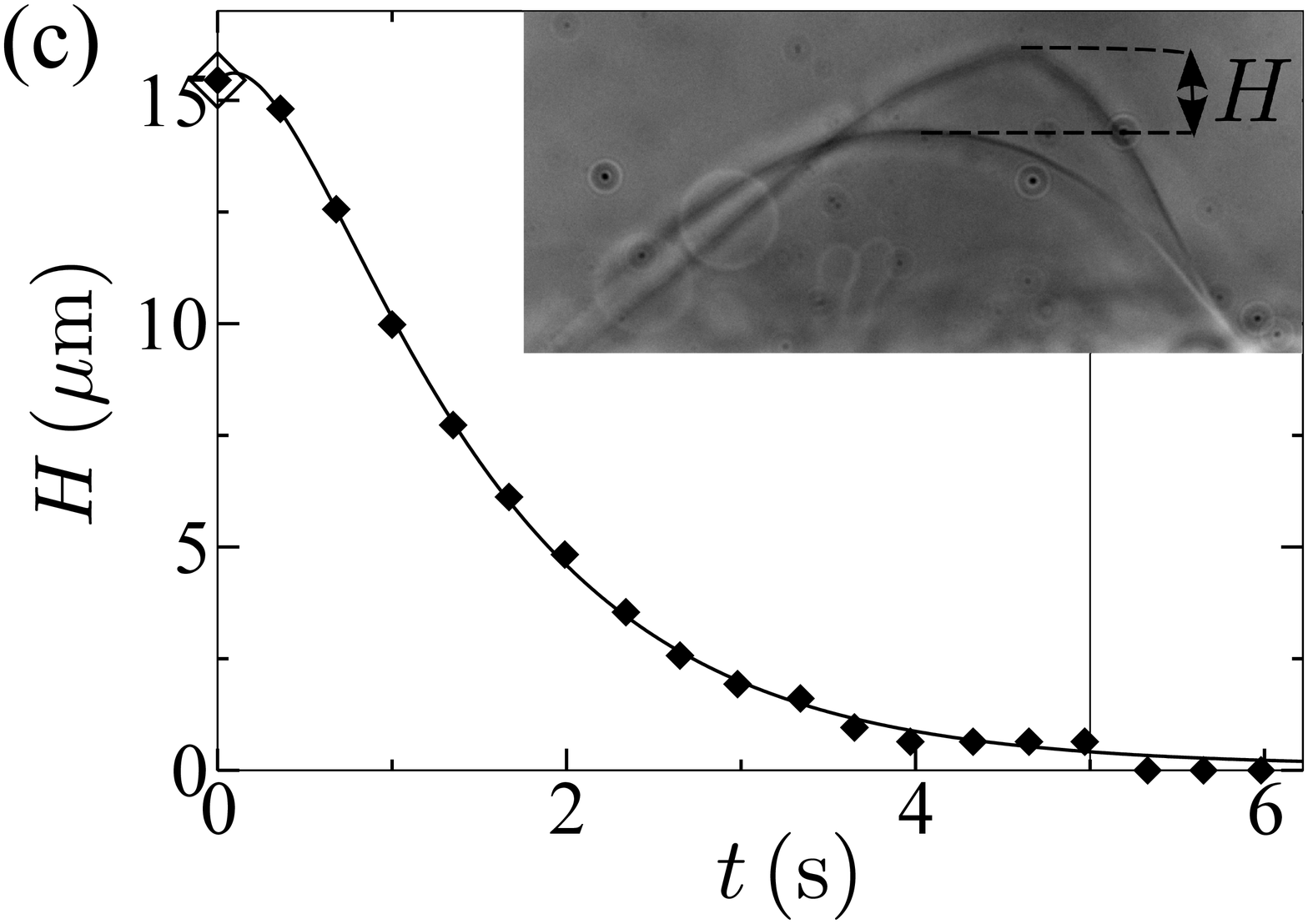}
\caption[]{Typical examples of fits of the experimental results. Dots: experimental data; line: best fit to Eq.~(\ref{Formule_fit}). The dots surrounded at $t=0$ correspond to the initial conditions. The lines at $t=5\,\mathrm{s}$ represent the upper bound of the time interval where the fit is carried out. The insets are superposed pictures of the initial shape of the GUV and of its most deformed shape in each experiment. The definition of $H(t)$ is recalled in these insets. (a) GUV 1 - Experiment number 5. (b) GUV 2 - Experiment number 8. (c) GUV 3 - Experiment number 15.\label{Exs_fits}}
\end{figure}

The best fits of the experimental data to Eq.~(\ref{Formule_fit}) are shown in Fig.~\ref{Exs_fits} for one typical ``pulse'' experiment for each of the three vesicles studied.
Fig.~\ref{Exs_fits} shows a good agreement between the experimental results and their best fit to Eq.~(\ref{Formule_fit}). Thus, this formula describes well the experimental results, which is in favor of our theoretical model. 

As can be seen on Fig.~\ref{Exs_fits}, the residual deformation at equilibrium is often close to zero. This residual deformation, which corresponds to the constant term $\kappa\bar c_0 \phi_q/(2\sigma_0)$ in Eq.~(\ref{evol_h_gen}), is due to the change of the spontaneous curvature induced by the chemical modification (see Sec.~\ref{Mod_subsec}).
Thus, if the diffusion of the hydroxide ions was much slower than the timescales of our instability, observing a negligible residual deformation would suggest that the change of the equilibrium density is predominant over the change of the spontaneous curvature. However, one cannot use this argument here, because the modified lipids may recover their initial state by reacquiring their protons as the hydroxide ions diffuse away in the solution (see Sec.~\ref{modif}). 

The best fits of the experimental results to Eq.~(\ref{evol_h_gen}) provide us with an estimate of the time constants $\gamma_1^{-1}$ and $\gamma_2^{-1}$, from which we may extract the values of the vesicle constitutive constants $b$ and $\sigma_0$, using Eq.~(\ref{eigenvalues}). 
However, the wavevector $q$, which comes from our single-mode description of the instability (see Sec.~\ref{hydro}), is involved in the expressions of $\gamma_1$ and $\gamma_2$  (see Eq.~(\ref{eigenvalues})). To calculate the characteristic wavevector $q$ involved in each experiment, we will measure the width $W$ of the instability, from which we will deduce an estimate of $q$ through $q\approx\pi/W$. 

\subsection{Estimation of the width of the instability}
\label{Width_subsec}
In order to estimate the characteristic width of the instability, we have measured the width at mid-height of the deformed zone of the vesicle. In practice, what we call ``deformed zone'' is the zone where the height of the vesicle is larger in the deformed state than in the initial state. 

\begin{figure}[h t b]
\centering
\includegraphics[width=0.45\textwidth]{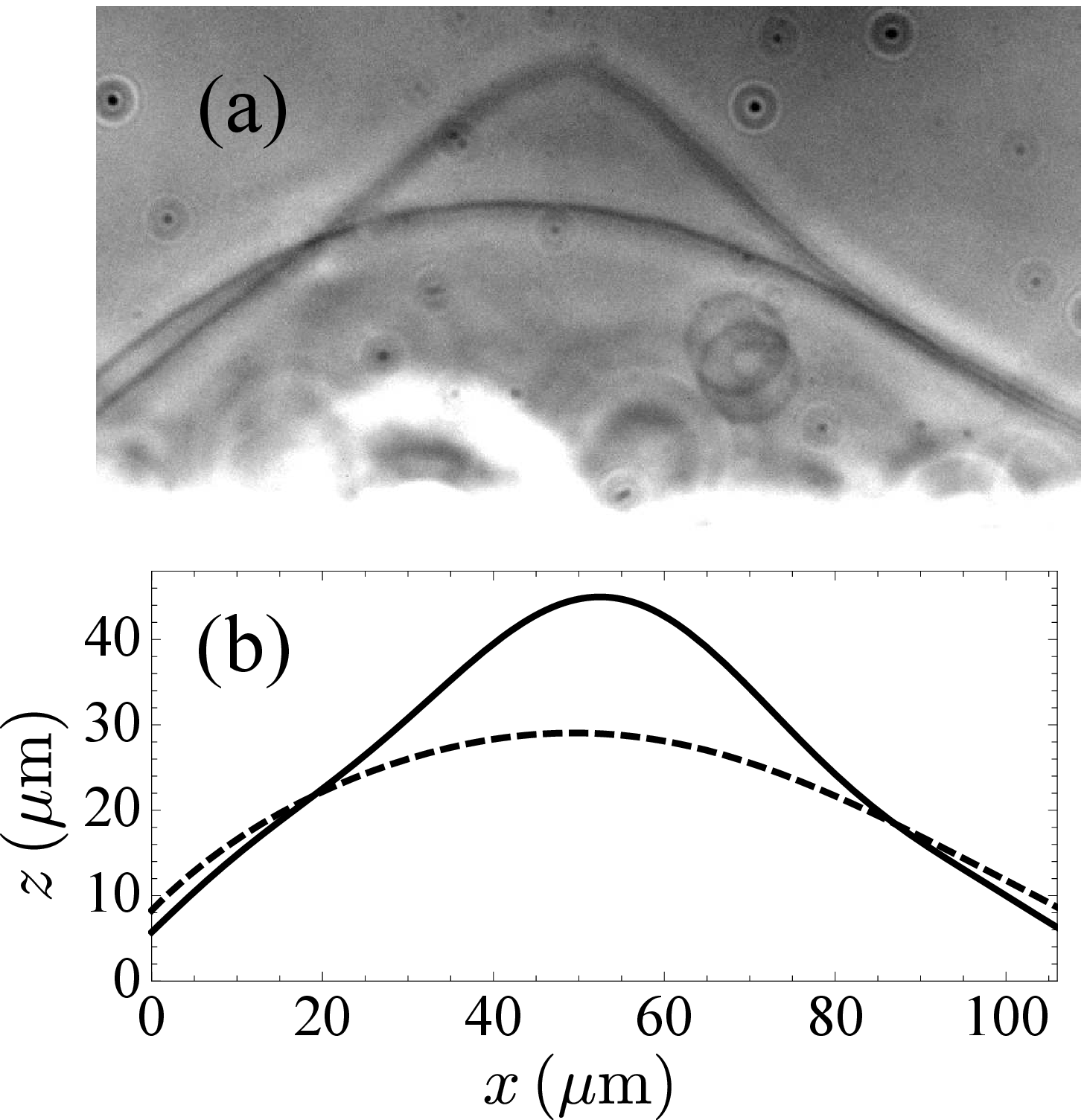}\\
\vspace{.2cm}
\includegraphics[width=0.45\textwidth]{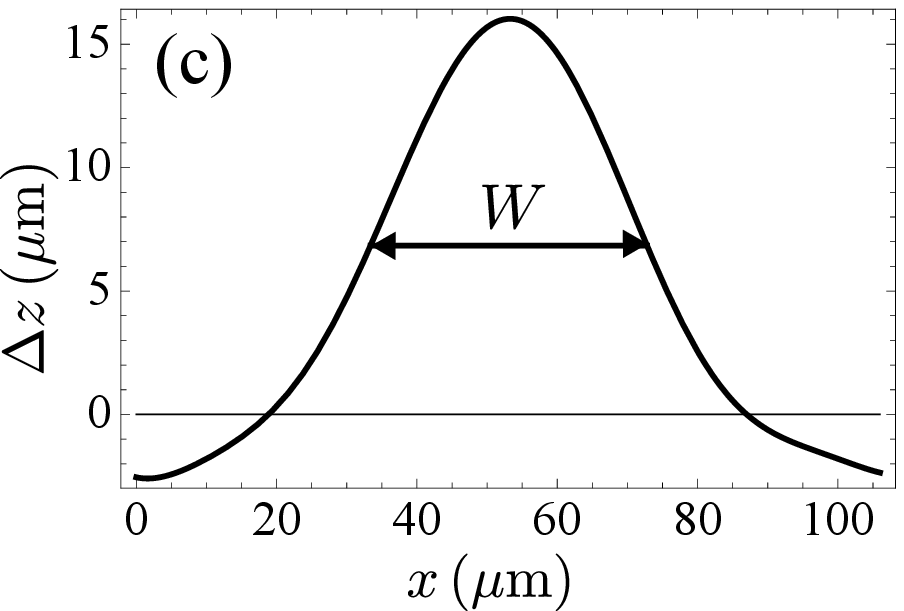}
\caption[]{(a) Superposed pictures of a GUV before the deformation and in the most deformed state (GUV 2, experiment number 6). (b) Polynomial fits of the two vesicle shapes in (a). (c) Plot of the difference between the deformed and the initial shape of the GUV. This difference is calculated by substracting the two fits in (b). It enables to determine the width at mid-height of the deformed zone, noted $W$.\label{width_fig}}
\end{figure}

We have digitized the profile of the vesicle before the approach of the pipette and in its most deformed state, i.e., when $H$ is maximal, which occurs just after the pipette is withdrawn. Fig.~\ref{width_fig}(a) shows the superposed pictures of a GUV in these two states during a typical ``pulse'' experiment. The digitized profiles at these two times have then been fitted to polynomials $z=P(x)$ (see Fig.~\ref{width_fig}(b)). The two fitting polynomials have then been substracted to get the deformation $\Delta z (x)$, from which it is straightforward to measure the width at mid-height of the deformed zone $W$ (see Fig.~\ref{width_fig}(c)).

This estimation of $W$ has been carried out on all of our ``pulse'' experiments. We thus have a specific estimate of the characteristic wavevector $q\approx\pi/W$ for each of these experiments. This value can be used to extract an estimate of the vesicle physical constants from the best fit of each experiment.

\subsection{Measurement of the intermonolayer friction coefficient $b$ and of the vesicle tension $\sigma_0$}

For each ``pulse'' experiment, the best fit of the experimental data to Eq.~(\ref{Formule_fit}), carried out following the method of Sec.~\ref{Fits_subsec}, provides an estimate of $\gamma_1$ and $\gamma_2$ (see Table~\ref{Tab}). Using the value of $q$ obtained for this experiment as explained in Sec.~\ref{Width_subsec} (see Table~\ref{Tab}), we may deduce the intermonolayer friction coefficient $b$ and the vesicle tension $\sigma_0$ from $\gamma_1$ and $\gamma_2$ thanks to Eq.~(\ref{eigenvalues}). 

\setlength{\tabcolsep}{0.15cm}
\begin{table}[h t b]
\footnotesize
\centering
\begin{tabular}{|l|lll|lll|lll|lll|lll|}
\hline
Exp.&\multicolumn{3}{c|}{$\gamma_1$ (s$^{-1}$)}&\multicolumn{3}{c|}{$\gamma_2$ (s$^{-1}$)}&\multicolumn{3}{c|}{$q$ ($\times 10^4$ m$^{-1}$)}\\
\hline
\hline
1&0.65&$\pm$&0.1&6.0&$\pm$&3&5.23&$\pm$&0.15\\
\hline
2&0.27&$\pm$&0.12&1.3&$\pm$&1.1&4.33&$\pm$&0.17\\
\hline
3&0.65&$\pm$&0.2&10&$\pm$&8&5.15&$\pm$&0.21\\
\hline
4&0.43&$\pm$&0.06&7.9&$\pm$&5&5.18&$\pm$&0.27\\
\hline
5&0.54&$\pm$&0.07&2.9&$\pm$&1.5&5.64&$\pm$&0.23\\
\hline
\hline
6&0.68&$\pm$&0.05&5.1&$\pm$&0.8&8.62&$\pm$&0.13\\
\hline
7&0.48&$\pm$&0.05&6.3&$\pm$&2&8.80&$\pm$&0.16\\
\hline
8&0.52&$\pm$&0.05&8.0&$\pm$&3&8.15&$\pm$&0.10\\
\hline
9&0.47&$\pm$&0.21&2.7&$\pm$&2&7.78&$\pm$&0.20\\
\hline
10&0.36&$\pm$&0.07&7.9&$\pm$&3&7.54&$\pm$&0.05\\
\hline
11&0.51&$\pm$&0.1&2.8&$\pm$&1.5&7.37&$\pm$&0.02\\
\hline
\hline
12&0.47&$\pm$&0.1&11&$\pm$&9&7.86&$\pm$&0.32\\
\hline
13&0.57&$\pm$&0.1&3.9&$\pm$&2&7.07&$\pm$&0.05\\
\hline
14&0.71&$\pm$&0.2&12&$\pm$&6&7.02&$\pm$&0.15\\
\hline
15&0.76&$\pm$&0.25&3.2&$\pm$&1.5&7.94&$\pm$&0.08\\
\hline
16&1.0&$\pm$&0.5&4.9&$\pm$&2&7.07&$\pm$&0.15\\
\hline
\end{tabular}
\normalsize
\caption[]{Values of the fitting parameters $\gamma_1$ and $\gamma_2$, and of the wavevector $q$ estimated as explained in Sec.~\ref{Width_subsec}, for each ``pulse'' experiment. The experiment numbers are the same as on Fig.~\ref{Fig_b}, and the thick horizontal lines indicate a change of vesicle. The intermonolayer friction coefficient $b$ and the vesicle tension $\sigma_0$ can be deduced from these values (see Fig.~\ref{Fig_b} for $b$). \label{Tab}}
\end{table}

\begin{figure}[h t b]
\centering
\includegraphics[width=0.55\textwidth]{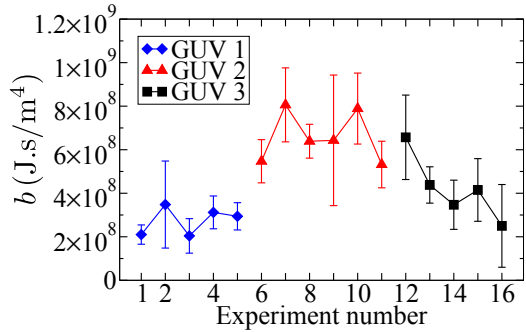}
\caption[]{Intermonolayer friction coefficient estimated from the best fit of the experimental data to Eq.~(\ref{Formule_fit}), for each ``pulse'' experiment.\label{Fig_b}}
\end{figure}

Fig.~\ref{Fig_b} shows the estimates of $b$ obtained from 16 different ``pulse'' experiments carried out on three different GUVs. The error bars correspond to the uncertainty on $b$ assuming that our model describes the experiment well. This uncertainty has several different origins. First, recall that we fit our experimental data on a $5\,\mathrm{s}$ time interval (see Sec.~\ref{Fits_subsec}), which is a somewhat arbitrary choice. In practice, the best fit, and therefore the estimated values of $\gamma_1$ (and $\gamma_2$), depend on the time interval over which the fit is carried out. This fact can be explained by the diffusion of the hydroxide ions, by the change in the equilibrium state which is sometimes observed (see Sec.~\ref{Fits_subsec}), and by the fact that the width of the deformation is in fact time-dependent (see Sec.~\ref{Sec_discussion}). We have thus varied the upper bound of this time interval from about 3 to 7 seconds, and taken the extremal values thus obtained for $\gamma_1$ (and $\gamma_2$) as the bounds of the uncertainty interval over $\gamma_1$ (and $\gamma_2$). This uncertainty interval is generally larger than the one estimated by the fitting software for a given time interval. Besides, estimating $b$ (and $\sigma_0$) also relies on our measurement of $W$. To determine the uncertainty regarding $W$, the measurement of $W$ described in Sec.~\ref{Width_subsec} has been carried out twice for each experiment, using the two most deformed states. All these factors have been taken into account in the error bars in Fig.~\ref{Fig_b}, the dominating one coming from the choice of the time interval.

The values we find for $b$, i.e., $b=2-8\times10^8\,\mathrm{J.s/m^4}$ (see Fig.~\ref{Fig_b}), are in good agreement with the literature \cite{Pott02, Shkulipa06}. Besides, we can see on Fig.~\ref{Fig_b} that the values of $b$ extracted from the different ``pulse'' experiments for each one of the three GUVs studied are compatible given the error bars. However, the uncertainty on $b$ is quite large, as shown by the error bars in Fig.~\ref{Fig_b}. Besides, our results seem to indicate that GUV 1 has a somewhat smaller intermonolayer coefficient friction $b$ than GUV 2, which was not expected, since all of our vesicles should have the same lipid composition, and $b$ should only depend on this composition. However, the existence of small composition variations in the experiments cannot be totally excluded. 

While the intermonolayer friction coefficient $b$ has been deduced from $\gamma_1$, which gives the longest timescale of the relaxation (a few seconds), the vesicle tension $\sigma_0$ has to be extracted from $\gamma_2$, which corresponds to a shorter timescale. The fits give values of $\gamma_2$ in the range $2-10 \,\mathrm{s^{-1}}$, so that the timescale $\gamma_2^{-1}$ is about $0.1$ to $0.5\,\mathrm{s}$ (see Table~\ref{Tab}). This is very short given that an experimental point is measured every $0.3\,\mathrm{s}$. Besides, studying Eq.~(\ref{ci_zero}) as a function of time shows that the term $e^{-\gamma_2 t}$ is of significant importance only before the maximum of $H(t)$ is reached, which confirms that data at very small times after the end of the injection would be needed to determine $\gamma_2$ well. Thus, it is impossible to extract precise values of $\sigma_0$ from our experiments. Even if more data points were available at short times, one might argue that the determination of $\sigma_0$ would still be imprecise because at short times, the pipette is still being withdrawn, and the resulting flow might affect the deformation.

We have nevertheless calculated estimates of $\sigma_0$ from each of our fits in the same way as $b$ was determined. We have obtained tensions in the range $0.6-6\times10^{-7}\,\mathrm{J/m^2}$, without notable difference between the three vesicles (the dispersion among different experiments on a single vesicle is similar to the one among all the experiments). The order of magnitude of $\sigma_0$ is correct, although no precise values can be extracted as mentioned above. It corresponds to floppy vesicles, which is in agreement with their qualitative observation. 

Thus, fitting our experimental results provides a satisfying order of magnitude for $b$ and $\sigma_0$, which is an argument in favor of our model. The measurements of $b$ are more precise than the ones of $\sigma_0$, due to the short timescale $\sigma_0$ is involved in. Besides, the observation of a relaxation with a timescale of a few seconds which is well-described by a term in $e^{-\gamma_1 t}$ shows that the change of the plane-shape equilibrium density is important in our instability (see Sec~\ref{comp}).

\section{Discussion}
\label{Sec_discussion}

Let us now discuss the possible improvements of our work. First of all, given that the timescale of the diffusion of the hydroxide ions is comparable to the timescale of the relaxation of the deformation (see Sec.~\ref{modif}), it would be very interesting to visualize the time-dependent pH field during the experiment, and even better to visualize directly the modified lipids thanks to specific fluorescent markers. This would settle whether \textit{(SH$_1$)} is verified or not, and for how long. Besides, the model could be improved by taking into account the diffusion of the hydroxide ions, so that \textit{(SH$_1$)} would no longer be necessary. However, it would be necessary to make strong assumptions on the way the hydroxide ions are injected and diffuse.  

\begin{figure}[h t b]
\centering
\includegraphics[width=0.45\textwidth]{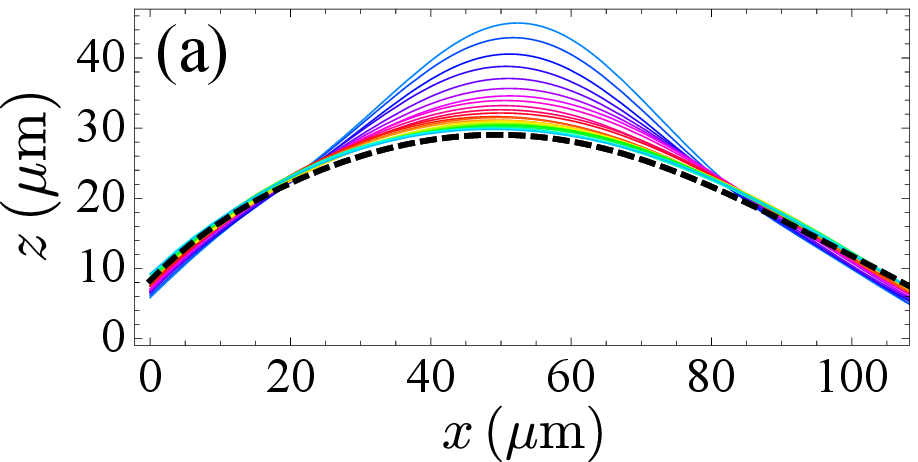}\\ \vspace{0.2cm}
\includegraphics[width=0.45\textwidth]{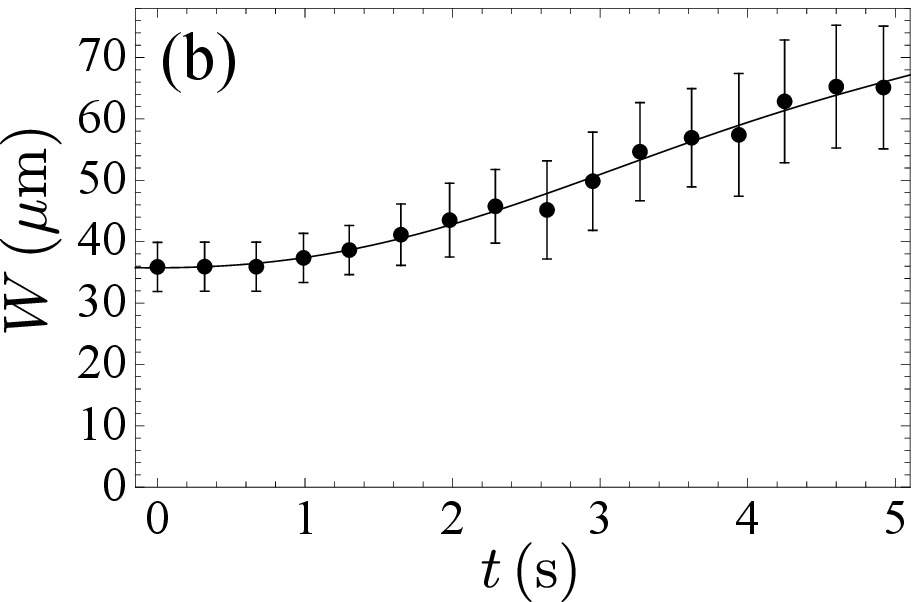}
\caption[]{(a) Time evolution of the vesicle shape during a typical ``pulse'' experiment (GUV 2, experiment number 6) after the pipette is withdrawn (bold dashed line: initial shape). (b) Width $W$ of the deformation as a function of time, extracted from the data in (a). \label{time_evol}}
\end{figure}

Another point of our model that could be improved lies in \textit{(SH$_2$)}: in practice, the deformation of the vesicle is not single-mode, so it would be an improvement to study the evolution of an initial deformation with a complete Fourier expansion. This is work in progress, but given the uncertainty on the spatio-temporal dependence of $\phi$ in the present experiments, it is difficult to gain further insight, and at the moment, the one-mode approximation is the best we can do. Besides, we have used the initial value of the width of the deformation $W$ to calculate the characteristic wavevector $q$. However, in reality, this width can change in time during the instability, due to dispersion, which does not occur for our theoretical single-mode deformation. In order to have an idea of the importance of this phenomenon, we have digitized the vesicle shape for all the experimental points in one ``pulse'' experiment, and calculated the width $W$ for each of these points, using the method in Sec.~\ref{Width_subsec}. The corresponding results are presented in Fig.~\ref{time_evol}.
We can see that $W$ is not constant during the instability. Its variation is about 90\% during the five second interval used for our fits. Nevertheless, it remains nearly constant just after the end of the injection before increasing: using the initial value of $W$ was thus the best thing we could do while keeping a constant $W$.

In order to determine $\sigma_0$ from these experiments, it would be useful to have more points at short times after the end of the injection. Another further experimental improvement would be to use larger vesicles, and to make the injection even more local than in the present work, so as to be closer to the assumptions of the model. 

\newpage

\section{Conclusion}
In this paper, we have reported experiments in which a local curvature instability is observed when GUVs are submitted to a local pH increase. 

A theoretical model of the dynamics of the instability has been described. We have shown that the chemical modification of the lipids resulted in a change of the spontaneous curvature and in a change of the plane-shape equilibrium density. Both of these effects have been taken into account and compared. In our description of the dynamics of the instability, the intermonolayer friction plays a crucial part: it gives the longest timescale of the instability, as the monolayers are unable to slide with respect to each other on short timescales.

Our model has been compared to the experiments by fitting the experimental data to the theoretical formula describing the height of the deformation as a function of time. The agreement between the experiments and the model is quite good. The intermonolayer friction coefficient can be extracted from these fits, yielding values consistent with the literature. Finally, we have discussed possible further developments of our work.

Studying the dynamics of our instability caused by a local chemical modification of the lipids enables to distinguish between a change of the plane-shape equilibrium density and a change of the spontaneous curvature. This is impossible in the case of a global modification of the environment of a vesicle. The fact that we observe a relaxation dynamics which is well-described by a mechanism involving the intermonolayer friction indicates that the change of the plane-shape equilibrium density is important in our instability.

\vspace{.5cm}
\bibliographystyle{unsrt}

\end{document}